\documentclass[aps,preprint,floatfix]{revtex4} 
\usepackage{amsmath}
\usepackage{amssymb}
\usepackage{amsfonts}
\usepackage[dvips]{graphicx}
\usepackage{epsfig}
\usepackage{tabularx}
\usepackage{color}
\usepackage{siunitx}
\usepackage{orcidlink}

\usepackage{lineno} 
\begin{document}
\title{Iterative map with power-law scaling of Gamma-distributed fluctuations related to prime numbers}

\author{Marzena Ciszak~\orcidlink{0000-0003-1087-6105}}
	\affiliation{CNR-INO Sede Secondaria di Sesto Fiorentino - c/o LENS, Via Nello Carrara, 1 - 50019 Sesto Fiorentino, Italy}

\date{\today}

\begin{abstract}
Prime numbers arise in several contexts beyond number theory, including statistical mechanics, quantum mechanics, and dynamical systems. However, the mechanisms underlying the irregularities of their sequence and their connections to physical systems remain poorly understood. The present work provides further insight into the search for deterministic fingerprints in the prime sequence. To this end, prime gaps at different separation distances are investigated through an empirical analysis. Based on this analysis, a modified local approximation of the Prime Number Theorem is introduced and analyzed empirically, in which the logarithmic gap estimate is evaluated at a midpoint-corrected argument. From this relation an iterative map is obtained that reproduces the classical asymptotic prime spacing at leading order and satisfies an approximate semigroup property. The residual fluctuations are found to follow Gamma distributions rescaled by the correction terms, with a variance exhibiting an approximate power-law scaling in the prime separation distance.
\end{abstract}


\maketitle

\section{Introduction}\label{sec1}
Prime numbers constitute an infinite subset of the natural numbers, defined by the elementary condition of divisibility only by unity and themselves. Despite this simple definition, their distribution is strikingly irregular and has fascinated scholars from antiquity to the present day. This apparent randomness has motivated sustained efforts to uncover a universal law governing the generation of successive primes \cite{cipolla1902,formulaCipolla,axler2017}.
Beyond their central role in number theory (see \cite{review,iwk} and references therein), prime numbers have increasingly attracted attention across the natural \cite{biology} and physical sciences \cite{physics}. Notably, they emerge in contexts spanning statistical mechanics, quantum mechanics and dynamical systems. A seminal study by Odlyzko \cite{odlyzko1987} identified a correspondence between the statistical distribution of the zeros of the Riemann zeta function and the eigenvalue spectra of random Hermitian matrices, within the framework of Random Matrix Theory originally developed in nuclear physics. Subsequent works have reinforced these connections, revealing parallels between prime numbers and quantum systems \cite{zyl2003,cipra1996,marek2014}. Prime numbers have been linked to quantum entanglement dynamics \cite{entanglement2023} and quantum models have been proposed that encode their arithmetic structure \cite{trombettoni2020,wolf1999}. Approaches inspired by quantum physics have also been applied to prime factorization exploiting logarithmic energy spectra \cite{logEnergy2018}. Complementary statistical-mechanical studies have uncovered unexpected correlations and emergent order in prime gaps distributions revealing their non-random structure \cite{stanley2000information,ares2006,correlations_zhang}, while connections to fractal organization \cite{wolf1997,wolf1989} and dynamical systems \cite{iovane2008,bonanno2004} have also been reported.  Interestingly, statistical analyses suggest the existence of periodicity and chaos in prime sequences \cite{bonanno2004,chaos,chaos2}. Collectively, these findings challenge the classical view of primes as purely random and indicate that prime numbers exhibit structural features reminiscent of complex phenomena observed in physical systems. However, the mechanisms responsible for these connections to certain physical systems remain largely unresolved, and it remains unclear to what extent the irregularities observed in the prime sequence reflect an underlying dynamical structure. One fruitful approach consists in examining prime gaps \cite{zhang,maynard,ford}. While much of the literature concerns nearest-neighbor prime gaps and their local statistical properties \cite{oliveira2014,cohen2024,marek2014,szpiro2004,scafetta2004}, less attention has been devoted to empirical analysis of the behavior of gaps across multiple separation scales.

The local approximation to the Prime Number Theorem can be used to estimate the expected number of primes in a short interval and, consequently, to obtain heuristic estimates for the typical size of prime gaps \cite{huxley1972,montgomery2007,lemke}. In this work, a modified form of the local approximation of the Prime Number Theorem is introduced and considered at various separation distances. This allows the construction of an iterative map for the leading-order growth of successive primes, which satisfies an approximate semigroup property. The fluctuations around this leading-order behavior are then analyzed. It is observed that the residual fluctuations are well described by a shifted and scaled Gamma distribution conditioned on primes, with the shape parameter following a power-law scaling asymptotically with the distance between primes. These results are captured by the proposed statistical model.

The remainder of the paper is organized as follows. First, a modified local approximation of the Prime Number Theorem is introduced, and the associated iterative map is derived and its properties analyzed. Then, the residual fluctuations are characterized and the statistical model is summarized. Finally, the conclusions are presented, followed by a discussion of the perspectives for future research. 

\begin{figure}[ht]
    \centering
    \includegraphics[width=1.\linewidth]{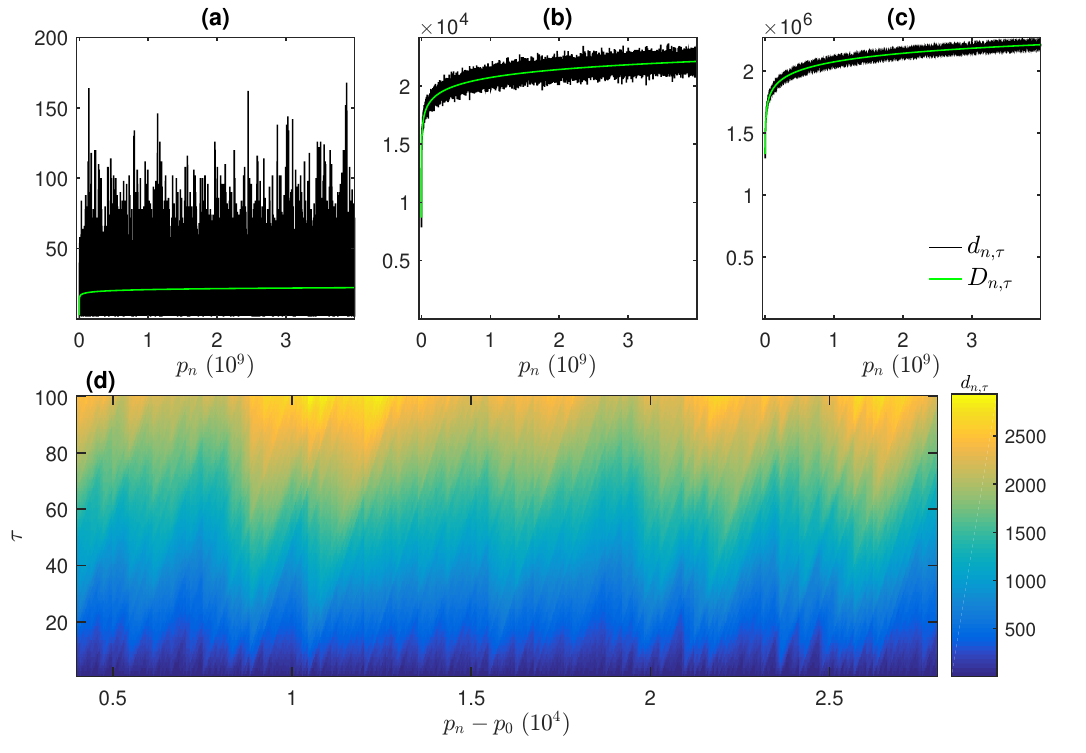} 
    \caption{Prime $\tau$-gaps $d_{n,\tau}$ for (a) $\tau=1 $,  (b) $\tau=10^3$ and (c) $\tau=10^5$. The data are shown under sampling every $10^4$ points. Green lines show the evolution of the iterative map (defined in Eq. \ref{eq:iterativemap}). (d) Prime $\tau$-gaps $d_{n,\tau}$ as a function of $\tau$. An offset $p_0=19999970000$ has been subtracted from $p_n$ for clarity.}
    \label{fig:1}
\end{figure}

\section{Results}

\subsection{Iterative map from leading growth of generalized prime gaps}\label{sec:composition}
Let's denote $p_n$ the successive prime numbers where $n=1,..,\pi(N)$ is the prime index and $\pi(N)$ is the prime counting function for primes less than or equal to $N$. Here, the study of prime gaps is extended to prime differences at different separation distance $\tau$ or generalized prime gaps (hereafter denoted by $\tau$-gaps): $d_{n,\tau}=p_{n+\tau}-p_n$ with $n\leq N_p$ for $N_p=\pi (N)-\tau$, where the shift index $\tau\geq1$ defines the length of separation between primes. 

In Fig. \ref{fig:1} (a) $\tau$-gaps between primes are plotted for $\tau=1$, while in Fig. \ref{fig:1} (b) and (c) $\tau$-gaps are plotted for $\tau=10^3$ and $\tau=10^5$, respectively. The average growth of prime gaps is $d_{n,\tau}\sim\tau \ln p_n$, according to the local Prime Number Theorem approximation \cite{montgomery2007,huxley1972}. Based on numerical observations, a slightly modified version is proposed and its properties are investigated empirically:
\begin{eqnarray}
  d_{n,\tau}\sim D_{n,\tau}=\tau \ln(p_n+c\tau)
   \label{eq:1} 
\end{eqnarray}
where $c$ is a free parameter. Numerical analysis indicates that incorporating a term $c\tau$, with $c>0$, provides a better approximation of the leading-order growth than the case $c=0$, as will be analyzed in detail in Section \ref{c}. 

Expanding the logarithmic term gives
$\ln(p_n+c\tau) = \ln p_n + \frac{c\tau}{p_n} + \mathcal{O}\!\left(\frac{(c\tau)^2}{p_n^2}\right)$. The leading-order term, $\ln p_n$, reproduces the average spacing between consecutive primes predicted by the Prime Number Theorem \cite{cipolla1902}: since the $n$ primes $p_1,\dots,p_n$ lie in $[1,p_n]$, the mean gap between them is $\frac{p_n}{\pi(p_n)}=\frac{p_n}{n}\sim\ln p_n$, equivalently, the density of primes near $p_n$ is approximately $1/\ln p_n$. This holds only on average: individual gaps $p_{n+1}-p_n$ fluctuate widely around $\ln p_n$. 

Equation \ref{eq:1} may be written as follows:
\begin{eqnarray}
p_{n+\tau} =  p_n+\tau \ln(p_n+c\tau)
   \label{eq:2} 
\end{eqnarray}
Let $F_\tau(p_n)=p_n+\tau\ln(p_n+c\tau)$, then at $\tau=1$ the following iterative map may be obtained:
\begin{eqnarray}
p_{n+1}=F_1(p_n)
   \label{eq:2a} 
\end{eqnarray}
The next step gives $p_{n+2}=F_1(p_{n+1})=F_1(F_1(p_n))$, etc. giving after $\tau$ iterations:
\begin{eqnarray}
p_{n+\tau} = F_1^{(\tau)}(p_n)
   \label{eq:2b} 
\end{eqnarray}
 which approximates the leading size of a successive prime $p_{n+\tau}$ after $\tau$ iterations starting from initial prime $p_n$. The iterated values of the above map are the approximation of the function defined in Eq. \ref{eq:2} (i.e $F_1^{(\tau)}\approx F_\tau$), up to an error that will be analyzed below.  
 
Using the iterative map in Eq. \ref{eq:2b}, the following expression for the $\tau$-gap is obtained:

\begin{equation}
p_{n+\tau}
=
p_n
+
\sum_{k=0}^{\tau-1}
\ln\!\left(p_{n+k}+c\right),
\label{eq:recursive}
\end{equation}
The error between the expressions, in Eq. \ref{eq:2} and Eq. \ref{eq:recursive} is defined as:

\begin{eqnarray}
E_\tau
&=&
\sum_{k=0}^{\tau-1}
\ln\!\left(p_{n+k}+c\right)
-
\tau
\ln\!\left(p_n+c\tau\right)
\end{eqnarray}
Considering that $\ln\!\left(p_{n+k}+c\right)=\ln\!\left(p_n+k\ln(p_n+ck)+c+c\tau-c\tau\right)=\ln(p_n+c\tau)+\ln\left(1+\tfrac{k\ln(p_n+ck)+c-c\tau}{p_n+c\tau}\right)$, where $p_{n+k}=p_n+k\ln(p_n+ck)$, one obtains, to first order,
\begin{eqnarray}
E_\tau
\approx
\sum_{k=0}^{\tau-1}
\frac{k\ln\!\left(p_n+ck\right)+c-c\tau}{p_n+c\tau}
\end{eqnarray}
The validity of this first-order expansion requires the argument of the logarithm to remain small, i.e.\ $\tau \ll p_n/\ln p_n$; this condition is assumed throughout the remainder of this derivation.

To estimate $E_\tau$, assume the approximation
\begin{equation}
p_{n+k}
\approx
p_n
+
k
\left(
\ln p_n
+
\frac{ck}{p_n}
\right),
\label{eq:secondorder}
\end{equation}
which incorporates the first nonlinear correction to the average prime gap.
Therefore,
\begin{equation}
E_\tau
\approx
\frac{1}{p_n+c\tau}
\left[
\ln p_n
\sum_{k=0}^{\tau-1}k
+
\frac{c}{p_n}
\sum_{k=0}^{\tau-1}k^2
+
\tau(c-c\tau)
\right].
\end{equation}
Using $\sum_{k=0}^{\tau-1}k=\frac{\tau(\tau-1)}{2}$ and $\sum_{k=0}^{\tau-1}k^2=\frac{\tau(\tau-1)(2\tau-1)}{6}$, the leading-order error becomes
\begin{equation}
E_\tau
\approx
\frac{\tau(\tau-1)}
{p_n+c\tau}
\left[
\frac{\ln p_n}{2}
-c
+
\frac{c(2\tau-1)}{6p_n}
\right].
\label{eq:leadingerror}
\end{equation}
Neglecting the higher-order contribution inside the brackets, the dominant term is minimized by choosing
\begin{equation}
c
=
\frac{1}{2}\ln p_n,
\label{eq:cchoice}
\end{equation}
which cancels the leading bracket term, $\ln p_n/2 - c = 0$, leaving only the subleading contribution. With this choice,
\begin{equation}
E_\tau
\approx
\frac{\tau(\tau-1)(2\tau-1)}{6\,p_n(p_n+c\tau)}\,c
=
\frac{\tau(\tau-1)(2\tau-1)\ln p_n}{12\,p_n(p_n+c\tau)},
\end{equation}
so that, under the assumed condition $\tau \ll p_n/\ln p_n$ (for which $c\tau = \tfrac12\tau\ln p_n \ll p_n$, consistent with the expansion above),
\begin{equation}
E_\tau
=
O\!\left(
\frac{\tau^3 \ln p_n}{p_n^2}
\right).
\label{eq:Etau}
\end{equation}

Equation \ref{eq:2} thus becomes

\begin{equation}
p_{n+\tau}=p_n+\tau
\ln\!\left(
p_n+\frac{\tau}{2}\ln p_n
\right).
\label{eq:secondorderapprox}
\end{equation}
Introducing the iterative variable $x_m=p_n$ for $m=n/\tau$, the corresponding iterative map is

\begin{equation}
x_{m+1}
=
x_m
+
\tau
\ln\!\left(
x_m+\frac{\tau}{2}\ln x_m
\right).
\label{eq:iterativemap}
\end{equation}

The minimization determines the value of $c$ such that the single logarithmic evaluation reproduces, as accurately as possible, the accumulated logarithmic evolution over the whole interval. Normally, these two expressions (in Eq. \ref{eq:2} and Eq. \ref{eq:recursive}) are not equivalent because the logarithm is nonlinear; however, the existence of a value of $c$ that minimizes this difference indicates that the nonlinear evolution admits a compact effective description. Interestingly, the obtained form of $c$ corresponds to half the average local prime spacing, evaluated at the approximate midpoint of the interval $[p_n, p_{n+\tau}]$. It does not alter the leading asymptotic relation, but provides a natural second-order refinement that can be tested through the behavior of the residual error.

The composition property of the map defined in Eq. \ref{eq:iterativemap} will be now analyzed. 
The evolution over two consecutive intervals $\tau_1$ and $\tau_2$ is described by the composition:
\begin{equation}
x_{m+2}
=
F_{\tau_2}
\left(
F_{\tau_1}(x_m)
\right).
\label{eq:composition}
\end{equation}
Let:
\begin{equation}
y=
F_{\tau_1}(x)
=
x+
\tau_1
\ln\left(
x+\frac{\tau_1}{2}\ln x
\right).
\label{eq:ydefinition}
\end{equation}
Applying the second iteration gives:
\begin{equation}
F_{\tau_2}(y)
=
y+
\tau_2
\ln\left(
y+\frac{\tau_2}{2}\ln y
\right).
\label{eq:secondstep}
\end{equation}
Substituting the expression for $y$, the exact composition law is obtained as:
\begin{equation}
\begin{aligned}
F_{\tau_2}\circ F_{\tau_1}(x)
=&
x+
\tau_1
\ln\left(
x+\frac{\tau_1}{2}\ln x
\right)
\\
&+
\tau_2
\ln
\left[
x+
\tau_1
\ln\left(
x+\frac{\tau_1}{2}\ln x
\right)
\right.
\\
&\left.
+\frac{\tau_2}{2}
\ln
\left(
x+
\tau_1
\ln\left(
x+\frac{\tau_1}{2}\ln x
\right)
\right)
\right].
\end{aligned}
\label{eq:exactcomposition}
\end{equation}
The corresponding single-step evolution over an interval $\tau_1+\tau_2$ is:
\begin{equation}
F_{\tau_1+\tau_2}(x)
=
x+
(\tau_1+\tau_2)
\ln
\left(
x+
\frac{\tau_1+\tau_2}{2}\ln x
\right).
\label{eq:singlejump}
\end{equation}

In general, the nonlinear map does not satisfy the exact semigroup property,
\begin{equation}
F_{\tau_2}\circ F_{\tau_1}(x)
\neq
F_{\tau_1+\tau_2}(x),
\label{eq:nosemigroup}
\end{equation}
because the logarithmic increment depends on the evolving state. Nevertheless, for large $x$ the map admits the asymptotic
expansion:
\begin{equation}
F_{\tau}(x)
=
x+\tau\ln x
+\frac{\tau^2\ln x}{2x}
-\frac{\tau^3(\ln x)^2}{8x^2}
+
O\left(
\frac{\tau^4(\ln x)^3}{x^3}
\right).
\end{equation}
Substituting this expansion into the composition
$F_{\tau_2}(F_{\tau_1}(x))$, one finds that all terms of order
$(\ln x)/x$ cancel with the corresponding terms in
$F_{\tau_1+\tau_2}(x)$. More precisely,
\begin{equation}
\begin{aligned}
F_{\tau_2}\circ F_{\tau_1}(x)
-
F_{\tau_1+\tau_2}(x)
&=
\frac{\tau_1\tau_2(\tau_1+\tau_2)}
{8x^2}
\left[
4\ln x-(\ln x)^2
\right]
\\
&\qquad
+
O\left(
\frac{(\ln x)^3}{x^3}
\right).
\end{aligned}
\end{equation}
Thus, although the map does not form an exact semigroup, it satisfies an approximate semigroup property to a higher asymptotic order:
\begin{equation}
F_{\tau_2}\circ F_{\tau_1}(x)
=
F_{\tau_1+\tau_2}(x)
+
O\left(
\frac{(\ln x)^2}{x^2}
\right),
\qquad x\to\infty.
\end{equation}
\begin{figure}[ht]
    \centering
    \includegraphics[width=1.\linewidth]{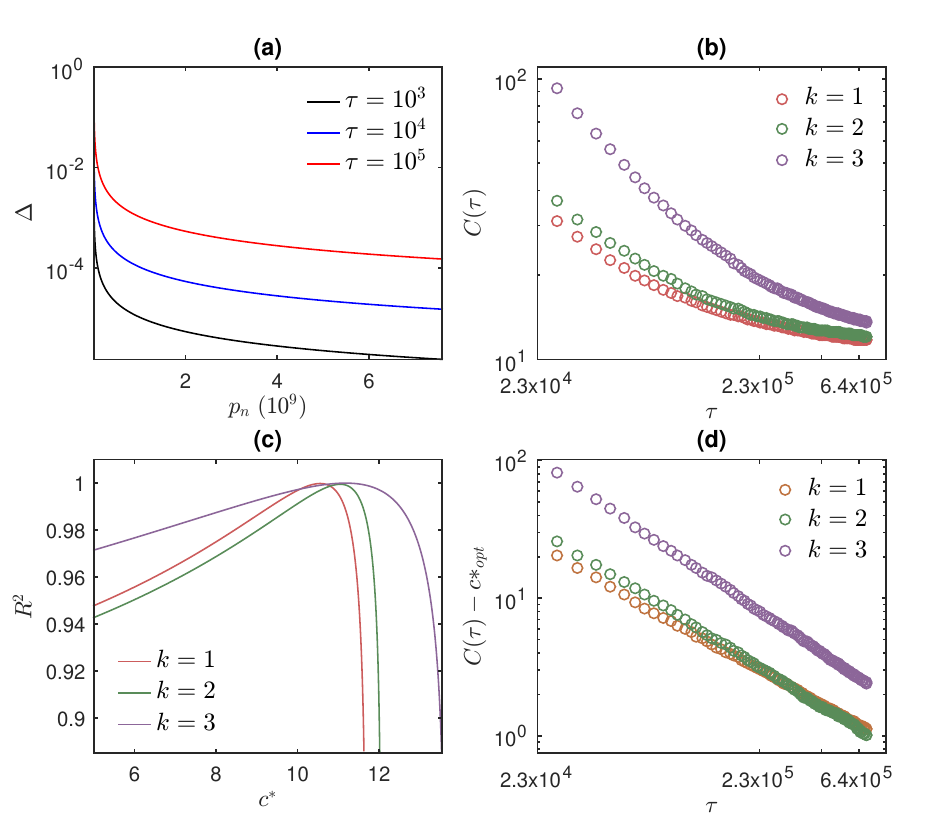}
    \caption{(a) Analysis of the effect of parameter $c$ on the term $\tau\ln(p_n+c\tau)$. The quantity $\Delta=\ln(p_n+\frac{1}{2}\ln p_n\tau)-\ln(p_n)$ is plotted for different values of $\tau$. (b) Empirically obtained values of $c$ minimizing $\bar{s}$ give a power-law function $C(\tau)$. $C(\tau)$ for different blocks of primes are shown. (c) The coefficient of determination $R^2$ versus $c^*$. (d) The data from inset (b) are plotted after subtracting $c^*_{opt}$.}
    \label{fig:2x}
\end{figure}
This small error represents the curvature of the logarithmic evolution: two consecutive prime-index intervals reproduce the effect of a single larger interval up to a small composition defect, arising from the nonlinear dependence of the increment on the current state. In the context of prime number approximations, this term measures the deviation from a purely additive model of prime gaps. It is worth noting that the inclusion of the midpoint correction $c=\frac{\tau}{2}\ln x$ inside the logarithm cancels the leading $O((\ln x)/x)$ composition defect that would otherwise be present for the map $F_\tau(x)=x+\tau\ln x$, corresponding to $c=0$.

The evolution of the iterative map defined in \ref{eq:iterativemap} is plotted in Fig. \ref{fig:1} (a-c) (in green color) for various values of parameter $\tau$. While in Fig. \ref{fig:1} (d) empirically calculated prime $\tau$-gaps are plotted as a function of $\tau$. 

The iterative sequence obtained from Eq. \ref{eq:iterativemap} is a smooth function describing the average growth of the prime $\tau$-gaps, however it does not account for the irregular fluctuations. These fluctuations will be the subject of the further analysis in Section \ref{sec:fluctuations}.

\subsection{Numerical estimation of $c$} \label{c}

As was previously shown, the specific choice of $c\equiv c(p_n)$ makes it possible to derive the iterative map consistent with the compact form of Eq. \ref{eq:2} up to an error $E_\tau$ (see Eq. \ref{eq:Etau}), which is smaller compared with the case $c=0$. To assess how well this specific form of $c$ holds, the mean $s$ is numerically analyzed for various constant values of $c$ over several ranges of prime magnitudes.
Accordingly, the prime sequence is partitioned into consecutive blocks $B_k$, each containing $L=10^8$ primes: $B_k = \left\{p _{(k-1)L+1}, \ldots, p_{kL}\right\}$. It should be noted that as $p_n\rightarrow \infty $ the term $c\tau$ becomes negligible with respect to $p_n$ and therefore the difference between both cases ($c=\frac{1}{2}\ln p_n$ vs $c=0$) decreases. This can be clearly seen in Fig. \ref{fig:2x} (a) where $\Delta =\ln (p_n+\frac{1}{2}\ln p_n\tau)-\ln (p_n)$ is plotted for various values of $\tau$. Considering this fact, the numerical analysis is therefore performed for the initial blocks with $k=1,2,3$, where the effect of parameter $c$ may be still easily captured from the numerical point of view. The results show that a function $C(\tau)$ emerges for each $k$-th block (see Fig. \ref{fig:2x} (b)) that is consistent with a power-law decay $C(\tau)=a^*\tau^{-b^*}+c^*$, where asymptotically $C(\tau)\rightarrow c^*$ as $\tau$ increases. In order to find the optimal value of $c^*$, the most linear curve in log-log space was estimated by performing a linear regression for various values of $c^*$ and identifying the value $c^* = c^*_{opt} \pm \Delta c^*$ that maximizes the coefficient of determination, $R^2$ (see Fig. \ref{fig:2x} (c-d)). 

Since the $R^2(c^*)$ curve is smooth, the tolerance $\delta R^2$ used to
define the parameter uncertainty $\Delta c^*$ was estimated from the classical asymptotic
approximation for the sampling variance of the coefficient of
determination, $\mathrm{Var}(R^2) \approx \frac{4\,R^2 (1-R^2)^2}{T}$,
where $T$ is the number of data points used to compute $R^2(c^*)$. The
tolerance was set to $\delta R^2 = \sqrt{\mathrm{Var}(R^2_{\max})}$,
evaluated at the fitted maximum $R^2_{\max}$, so that $\Delta c^*$
reflects the range over which $c^*$ cannot be statistically distinguished
from $c^*_{opt}$ given the finite size of the underlying dataset.
The optimal values of the parameter $c^*$ for the considered blocks $k=1,2,3$ of primes were found to be 
\[(c^*_{opt}\pm\Delta c^*)_k=(10.56\pm0.09,11.05\pm0.12,11.18\pm0.12),\] 
corresponding to a maximum coefficient of determination $(R^2_{\max})_k = (0.9997,0.9994,0.9999)$. 
On the other side, the mean values of $c=\tfrac{1}{2}\ln p_n$ calculated over a given block $k$ as follows, $(\bar{c})_k=\tfrac{1}{2L}\sum_{j=(k-1)L+1}^{kL}\ln p_j$, are
\[(\bar{c})_{k}=(10.69,11.08,11.29)\].   
which may be compared to those obtained for $c^*_{opt}$.

The two different types of analysis, numerical to minimize $\bar{s}$ and theoretical derivation of the iterative map from the local Prime Number Theorem approximation, indicate similar values for the parameter $c$.   

\subsection{Statistical analysis of residuals} \label{sec:fluctuations}

\begin{figure}[ht]
    \centering
    \includegraphics[width=1.\linewidth]{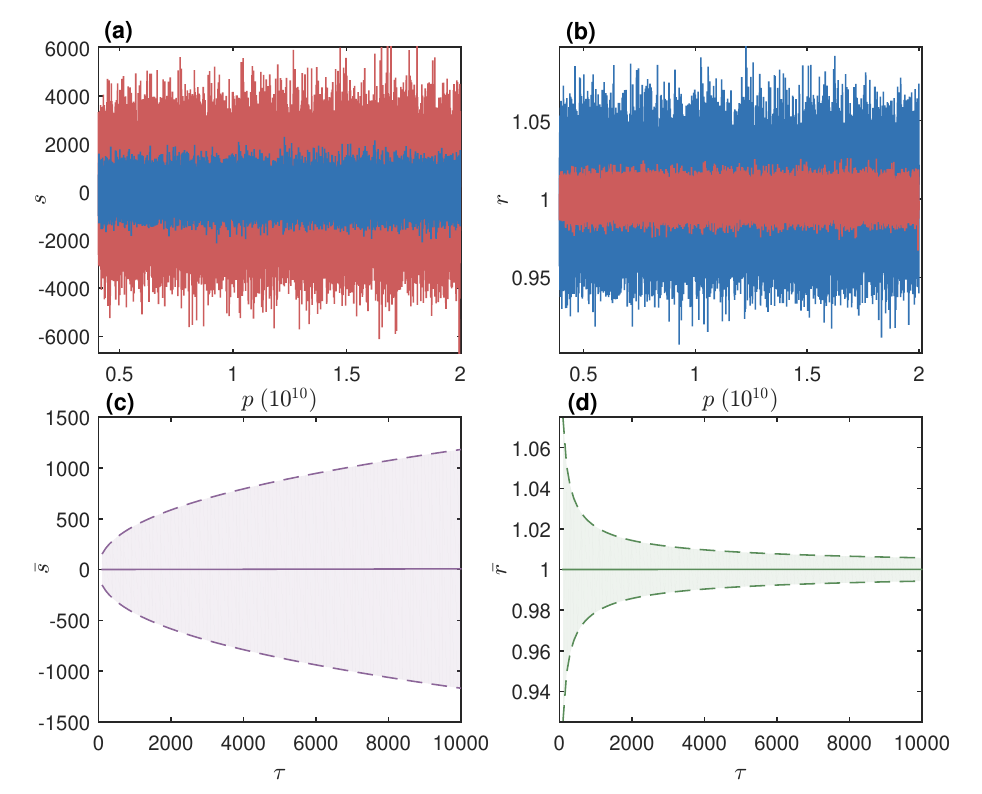}
    \caption{Subtractive residuals $s$ (a) and relative residuals $r$ (b) for $\tau=10^3$ (blue line) and $\tau=10^4$ (brown line). One point is plotted every $10^3$
 data points. (c) Representative average subtractive residual $\bar{s}$ over evolving variability in $p_n$ calculated for fixed interval and (d) average relative residuals $\bar{r}$ (solid lines) with corresponding standard deviations (dashed lines) as a function of $\tau$. }
    \label{fig:3}
\end{figure}

Using Eq. \ref{eq:1}, the subtractive residuals are defined for each $p_n$ and $p_{n+\tau}$:
\begin{eqnarray}
s = d_{n,\tau}-D_{n,\tau} 
   \label{eq:4} 
\end{eqnarray}
giving a set of $N_p$ values of $s$ with approximately zero mean and with a non-stationary variance \(\sigma_S^2\) (see Fig. \ref{fig:3} (a)). This variance increases both with $p_n$ as well as with $\tau$ that can be seen in Fig. \ref{fig:3} (c), where the representative standard deviation over evolving variability is shown. 

The relative residuals may be also defined:
\begin{eqnarray}
r=\frac{d_{n,\tau}}{D_{n,\tau}}=\frac{s}{D_{n,\tau}}+1
\label{eq:5}
\end{eqnarray}
that gives a set of $N_p$ values of $r$ with a mean $\bar{r}\approx 1$ and variance \(\sigma_R^2\) (see Fig. \ref{fig:3} (b) and (d)). In this case, the variance of $r$ turns out to be approximately stationary in $p_n$, what allows the distributions to be analyzed.
Moreover, let $x_r^{(1)}, x_r^{(2)}, \dots, x_r^{(N_p)}$ be independent and identically distributed random variables drawn uniformly from the interval $[2, N]$, where $N$ is the upper bound considered in the prime analysis. Let $x_r(1), x_r(2), \dots, x_r(N_p)$ denote the corresponding order statistics such that $x_r(1) \le x_r(2) \le \dots \le x_r(N_p)$. We define the $\tau$-spacings of the ordered sample by $\Delta_\tau(n) = x_r(n+\tau) - x_r(n)$ for $n = 1, \dots, N_p - \tau$. To compare with prime $\tau$-gaps, we introduce the normalized increment $r_{\mathrm{rand}}(n) = \frac{\Delta_\tau(n)}{\bar{\Delta}_\tau }$, where $\bar{\Delta}_\tau $ denotes the average over $n$.
This construction defines a reference stochastic process based on spacings of ordered random samples, which can be interpreted as a one-dimensional point process with non-negative increments. The resulting cumulative sums correspond to a renewal-type process, which under standard approximations yields Gamma-like statistics for aggregated displacements. This random model serves as a baseline for comparison with the structure of prime $\tau$-gaps.

\begin{figure}[ht]
    \centering
    \includegraphics[width=0.8\linewidth]{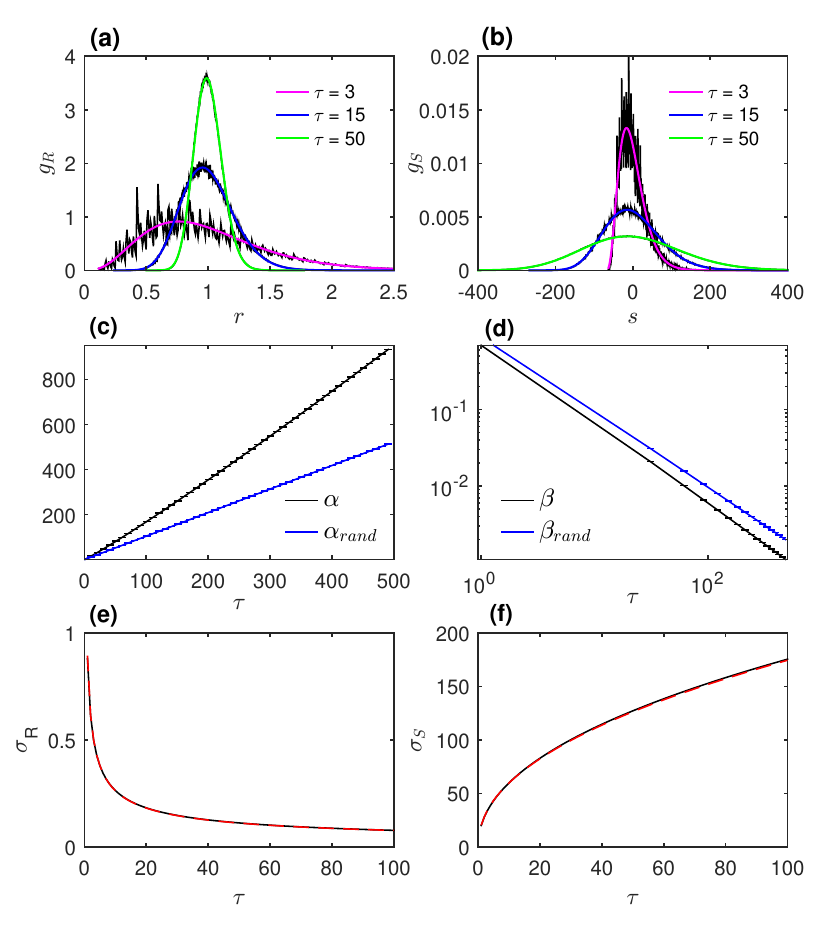}
    \caption{Distribution for $r$ (a) and $s$ (b) for different values of $\tau$ (black lines). The color lines show the model Gamma distributions with corresponding value of $\alpha(\tau)\sim\tau^{1+\gamma}$. Fitted Gamma distribution parameters (c) $\alpha$ for primes and $\alpha_{rand}$ for random case and (c) $\beta$ for primes and $\beta_{rand}$ for random case. The error bars represent the parameter uncertainties estimated from the maximum-likelihood distribution fitting. Standard deviations obtained numerically (solid black lines) and the predicted ones (dashed red lines) for (e) the relative residuals with $\sigma_R\sim\tau^{-\tfrac{1}{2}(1+\gamma)}$ and (f) subtractive residuals with $\sigma _S$ (calculated from Eq. \ref{eq:9} for $M=N_p/10^4$ elements). }
    \label{fig:4}
\end{figure}

The distributions of relative residuals $r$ for prime numbers, as in the case of random numbers, are found to be best fitted to a Gamma distribution $g_R(r)=\text{Gamma}(r; \alpha, \beta)$ with $\beta=1/\alpha$:
\begin{equation}
g_R(r) = \frac{\alpha^\alpha}{\Gamma(\alpha)} r^{\alpha - 1} e^{-\alpha r}, \quad r > 0, \; \alpha > 0
\label{eq:6}
\end{equation}
where \(R\) denotes a discrete variable with $\bar{R} =1$, and $\alpha$ and $\beta$ are a shape and scale parameters, respectively. In both, random and prime number cases, the shape parameter $\alpha$ depends on $\tau$, but the difference lays in the form of this dependence. In the case of the random numbers the shape parameter $\alpha _{rand}(\tau)\sim\tau$, while in the case of prime numbers it is nonlinear and follows the power law $\alpha(\tau)=a\tau^{1+\gamma}+b$ with $a=1.15\pm 0.001$, $\gamma= 0.081\pm 0.0001$ and $b=1.06 \pm 0.04$ (for primes in the range $[2,2\times10^{10}]$). Therefore, the resulting variance of the distribution is $\sigma_R^2=1/\alpha\sim \tau^{-(1+\gamma)}$. In Fig. \ref{fig:4} (a) the distributions for $r$ and Gamma distributions with respective $\alpha$ are shown for different values of $\tau$. It is worth noting that for $\tau=1$ the distribution reduces to an exponential one, which is consistent with reports in the literature \cite{marek2014,scafetta2004}. The dependence of $\alpha$ and $\beta$ on $\tau$ is displayed in Fig. \ref{fig:4} (c) and (d), respectively. The uncertainties in the fitted values of $\alpha$ and $\beta$, obtained from the maximum-likelihood distribution estimation, are nearly indistinguishable on the scale of the figure due to their small magnitude. In Figure \ref{fig:4} (e) the standard deviations obtained numerically and the predicted  ones for the relative residuals are shown.

For values of $\tau\gtrsim 50$ the distribution of relative residuals tends to normal distribution $\mathcal{N}(1,\sigma_{R})$. When $\tau \rightarrow \infty$ it collapses to a Dirac function $\delta (r-1)$ and the sequence of $\tau$-gaps follows the trend asymptotically, since the amplitude of fluctuating residuals goes to zero. This is caused by the fact that the division of two large, closely matched quantities suppresses information about fluctuations, as their ratio asymptotically approaches unity, thereby effectively filtering out the underlying variability. The situation changes however for the subtractive residuals, for which the variance clearly increases with $\tau$ as seen in Fig. \ref{fig:3} (c). The resulting Gamma distribution  for $r$ can be used to determine the distribution for $s$ since $r=\frac{s}{D_{n,\tau} }+1$. Then the change of variable results in $ g_S(s; \tau) = \left| \frac{dr}{ds}\right| g_R\left (\frac{s}{D_{n,\tau}}+1; \tau \right)=\frac{1}{|D_{n,\tau}|} g_R\left(s;\tau|p \right) $ and the following distribution for the subtractive residuals is obtained:
\begin{eqnarray}
 g_S(s; \tau|p)= \frac{1}{\left |D_{n,\tau}\right|}\frac{\alpha^\alpha}{\Gamma(\alpha)} \left(\frac{s}{D_{n,\tau}}+1\right)^{\alpha - 1} e^{-\alpha \left(\frac{s}{D_{n,\tau}}+1\right)}, \qquad s>-D_{n,\tau}
\label{eq:7} 
\end{eqnarray}
which is a shifted and scaled Gamma distribution conditioned on $p$ where \(S\) denotes a discrete variable with $\bar{S}(p) =0$ and variance $\sigma_S^2(p) = (D_{n,\tau})^2\sigma_R^2$. Since $D_{n,\tau}\equiv D_{n,\tau}(p)$ depends explicitly on $p$ the distribution of $S(p)$ varies deterministically with the prime number $p$. Consequently, $S(p)$ constitutes a non-stationary family of shifted and scaled Gamma-distributed random variables indexed by primes. 

Averaging over the primes $p$, the marginal (unconditional) distribution is obtained as:
\begin{eqnarray} 
 g_S(s;\tau)= \sum_{m=1}^{M-1} W_m g_S(s; \tau| p_m)
\label{eq:8} 
\end{eqnarray}
where $W_m=(p_{m+1}-p_m)/(p_M-p_1)$ are normalized weights proportional to the local prime gaps satisfying $\sum_{m=1}^{M-1}W_m=1$, since $\sum_{m=1}^{M-1}(p_{m+1}-p_m)=p_M-p_1$, where $M$ denotes the number of primes in the considered interval. The quantity $W_m$ may be interpreted as the relative contribution of the gap 
$p_{m+1}-p_m$ to the total interval $[p_1,p_M]$. The average size of the $m$-th prime gap satisfies $p_{m+1}-p_m\approx\ln p_m$,
hence:
\begin{equation}
W_m
\approx
\frac{\ln p_m}{\sum_{j=1}^{M-1}\ln p_j}
\label{eq:Wm}
\end{equation}
Suppose the primes are taken from an interval where the local density changes little. Then $\ln p_j \approx L$ for all $j=1,...,M-1$, where $L$ is some representative value. Then $\sum_{j=1}^{M-1}\ln p_j=(M-1)L$ and one gets 
\begin{equation}
    W_m=\frac{L}{(M-1)L}=\frac{1}{M-1}
\end{equation}
Thus, although individual prime gaps fluctuate, each normalized gap contributes, on average, approximately $1/(M-1)$ of the total interval.

Since the residual fluctuations $R$ have (approximately) zero conditional mean for every $p$, the between-component contribution to the total variance vanishes, and the law of total variance reduces to
\begin{eqnarray} 
\sigma_S^2 = \sigma_R^2 \sum_{m=1}^{M-1} W_m D^2_{m,\tau}
\approx  \tfrac{\tau^{1-\gamma}}{a} \tfrac{1}{M-1}\sum_{m=1}^{M-1} \Big(\ln(p_m+c\tau)\Big)^2
\label{eq:9} 
\end{eqnarray}
where $ D_{m,\tau}=\tau \ln(p_m+c\tau)$ was considered.
This shows that the variance is modulated by a slowly varying logarithmic correction arising from a nonlinear function of primes over the interval $[p_1,p_M]$ and exhibits an overall power-law scaling $\tau^{1-\gamma}$. It differs substantially from $\sigma_R^2$, which remains roughly unchanged regardless of the number of primes considered. 

Note that even in the case of a coarse-grained sampling $m = 1 + i \Delta m$, with $i = 0, 1, \dots, \left\lfloor \frac{M-1}{\Delta m} \right\rfloor$ and step size $\Delta m\gg1$, the weights remain approximately uniform, $W_m\approx\frac{1}{M/\Delta m}$. The resulting theoretical prediction for $\sigma_S^2$, describing the amplitude of the global fluctuations of the ensemble of prime gaps, shows good agreement with the numerical results (see Fig.~\ref{fig:4}(f)).

The observed variance form can be interpreted as a consequence of the mapping, provided the relative fluctuations of the residuals possess an intrinsic variance $\sigma_R^2=1/\alpha$ scaling as $\tau^{-(1+\gamma)}$. Each component $D_{m,\tau}$ acts, for a given state, as a multiplicative scale on these fluctuations, so its square multiplies the corresponding variance term in $\sigma_S^2$. Since $D_{m,\tau}\propto\tau$ at leading order, the variance of the residual scales as $\sigma_S^2\sim\tau^{1-\gamma}$.

\begin{figure}[ht]
    \centering
    \includegraphics[width=0.8\linewidth]{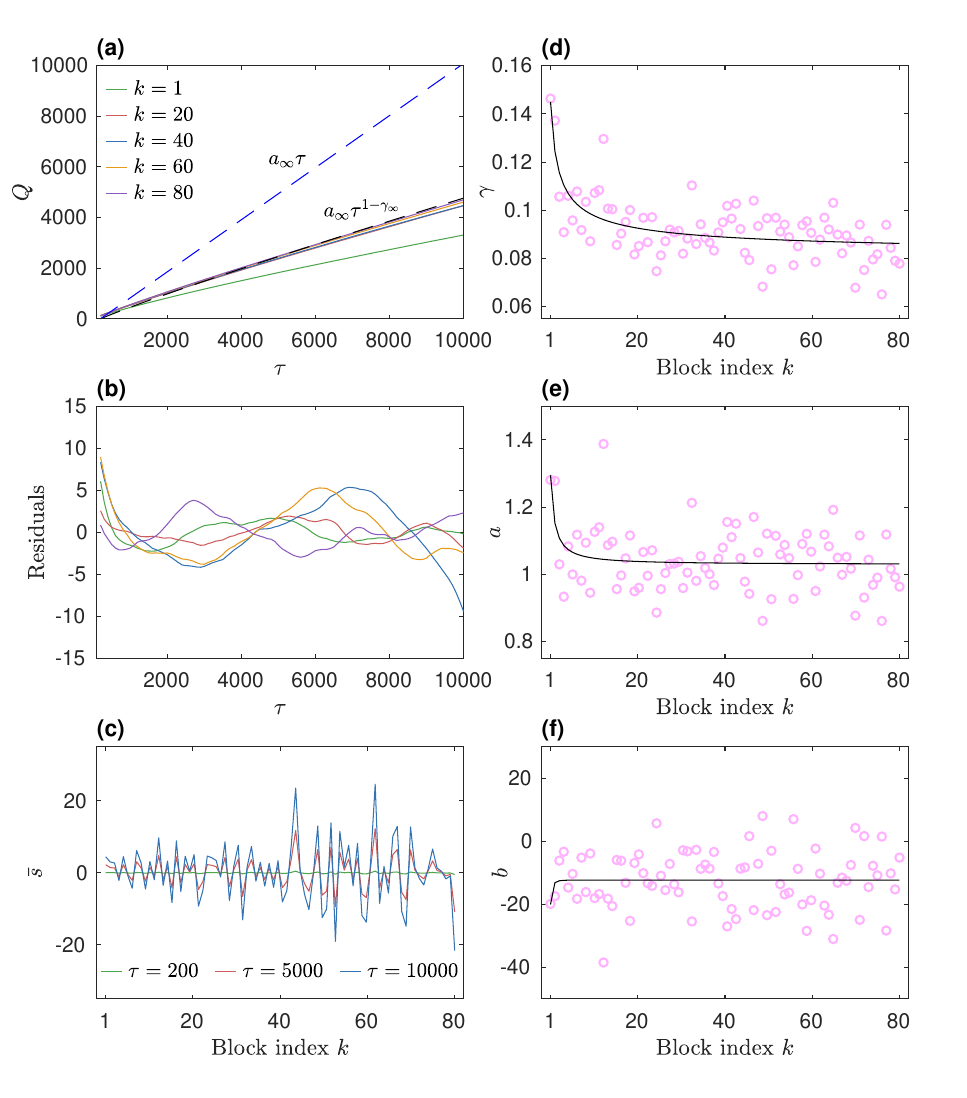}
    \caption{(a) The quantity $S$ as a function of $\tau$ for selected prime intervals. For comparison with the estimated by fitting $a_\infty\tau^{1-\gamma_\infty}$ (dashed black line), $a_\infty\tau$ is also plotted (dashed blue line). (b) Residuals from the power-law fit for the selected prime intervals (see legend in (a)). (c) Mean values of the subtractive residuals $s$ as a function of block index $k$ for selected values of $\tau$. Parameters (d) $\gamma$, (e) $a$ and (f) $b$ obtained from the power-law fit to the quantity $Q(\tau)=a\tau^{1-\gamma}+b$ as a function of block index $k$ (magenta circles). The black lines correspond to the power-law fits of $\gamma (k)$, $a(k)$ and $b(k)$, shown respectively in (d-f).}
    \label{fig:5}
\end{figure}

\subsection{Test of the variance scaling}
A direct test of the proposed variance relation can be obtained by separating the contribution induced by the logarithmic mapping from the intrinsic fluctuation of the residuals. The average squared value of the logarithmic scale factor associated with the mapping is:
\begin{equation}
A(\tau)=\sum_{m=1}^{M-1}W_m\ln\left(p_m+\tfrac{\tau}{2}\ln p_m\right)^2
\label{eq:average_scale_test}
\end{equation}
where $W_m$ is defined in Eq. \ref{eq:Wm}.
The quantity
\begin{equation}
Q(\tau)=
\frac{\sigma_S^2(\tau)}{A(\tau)}
\label{eq:Q_test}
\end{equation}
removes the contribution to the variance arising from the logarithmic scale of the mapping. If the proposed interpretation is correct, the remaining fluctuation should follow the intrinsic scaling
\begin{equation}
Q(\tau)\propto \tau^{1-\gamma}.
\label{eq:Q_scaling}
\end{equation}

The analysis is performed over several ranges of prime magnitudes in order to test the robustness of the predicted scaling. To this end, the prime sequence is partitioned into consecutive blocks, each containing $L=10^8$ primes. For the $k$-th block, $ B_k = \left\{p_{(k-1)L+1}, \ldots, p_{kL}\right\}$, the corresponding local statistical value of $\gamma$ is calculated. Thus, one value $\gamma(k)$ is obtained for each block, and the resulting block-wise statistics are analyzed as a function of the prime block index $k$.

In Fig.~\ref{fig:5}(a), the quantity $Q$ is plotted as a function of $\tau$ in the interval $\tau \in [200, 10^4]$, for selected values of the block index $k$. Local fits of the form $a\tau^{1-\gamma} + b$ for $Q(\tau)$, performed using the nonlinear least-squares method, yield residuals between the empirically obtained and estimated values, which are shown in Fig.~\ref{fig:5} (b). It is important to notice that the mean values $\bar{s}$ of the subtractive residuals oscillate irregularly around zero across the blocks $k$, ensuring the absence of drifts caused by finite-size effects (see Fig.~\ref{fig:5} (c)). Figure \ref{fig:5} (d) reveals that parameter $\gamma $ changes over $k$ and, in turn, is best fitted by a power-law dependence saturating to a fixed, non-zero value. The same situations occurs for parameters $a$ (see Fig. \ref{fig:5} (e)), and $b$ (see Fig. \ref{fig:5} (f)). More precisely, the obtained values, using the nonlinear least squares method assuming $k\rightarrow\infty $ give: 
\[(\gamma_\infty, a_\infty,b_\infty)=(0.08\pm 0.01,1.03\pm0.03,-12.26\pm2.12)\]. 
Therefore empirical analysis of data over prime blocks favor the interpretation that $\gamma$ approaches a positive constant and is consistent with power-law scaling found before. Nevertheless, because the measured range terminates at $k=80$ (corresponding to approximately $8\times 10^9$ primes in the interval $[2,2\times 10^{11}]$), this value should currently be regarded only as an indicative extrapolation suggesting $\gamma >0$ rather than a definitive measurement of the asymptotic limit.

\subsection{Statistical model summary}

The full picture developed in this work can be summarized by combining
the deterministic leading-order growth with the fluctuating residual
component in the following iterative map with the parameter $\tau$:
\begin{equation}
p_{n+\tau} = p_n + D_{n,\tau} + s,
\label{eq:full_model}
\end{equation}
where $p_n$ is the current prime,
$D_{n,\tau}=\tau\ln\!\left(p_n+\tfrac{\tau}{2}\ln p_n\right)$ is the
deterministic, separation-dependent leading-order gap, and $s$ is a
zero-mean stochastic residual. Given the state $p_n$,
$s=D_{n,\tau}(r-1)$, where $r$ is a unit-mean Gamma-distributed factor
with shape parameter $\alpha$ and variance $\sigma_R^2=1/\alpha$, so
that the distribution of $s$ is a weighted mixture of shifted and
rescaled Gamma distributions with weights $W_m$. To a good
approximation, the variance of $s$ is
$\sigma_S^2=\sigma_R^2\sum_{m=1}^{M-1} W_m D^2_{m,\tau}$, i.e., it
scales with the weighted squared correction terms, while $\sigma_R^2$
follows an approximate power law in the separation distance,
$\sigma_R^2\sim\tau^{-(1+\gamma)}$, so that the shape parameter scales
as $\alpha\sim\tau^{1+\gamma}$. Since $D_{m,\tau}\propto\tau$ at leading
order, the residual variance scales as $\sigma_S^2\sim\tau^{1-\gamma}$,
up to logarithmic factors.

The power-law scaling of the relative variance of the Gamma-distributed
residual factor indicates that the fluctuations are not characterized by a single, fixed
relative noise level, but show a scale-free dependence on the separation
$\tau$ \cite{mandelbrot1982fractal}. For independent gaps the relative
variance of their sum decreases as $\tau^{-1}$ (i.e., $\gamma=0$ for the random case), so
deviations from this value point to statistical dependence between
successive gaps \cite{beran1994longrange}. 

The exponent $\gamma$ thus characterizes the scaling regime of the fluctuations, in qualitative analogy with the Hurst-exponent classification of persistent and anti-persistent processes
\cite{hurst1951longterm}. The fitted exponent $\gamma\simeq0.08$ corresponds to a residual variance $\sigma_S^2\sim\tau^{0.92}$ (see $\sigma_S$ plotted in function of $\tau$ in Fig. \ref{fig:3} (c)), i.e., a Hurst-like exponent $H=(1-\gamma)/2\simeq0.46$, close to the uncorrelated
value $H=0.5$ and slightly on the anti-persistent side. Since the fluctuations are measured around the deterministic growth $D_{n,\tau}$, the non-stationarity of the mean gap is removed (see Fig. \ref{fig:5} (c)), which Ref.~\cite{scafetta2004} identified as the origin of the apparent persistence in the raw gap series. The small deviation from $H=0.5$ may therefore indicate a weak anti-persistent character of the
residual fluctuations. 

The consistency of the model \ref{eq:full_model} with composition (the semigroup property) can be examined by
splitting a jump of size $\tau_1+\tau_2$ into two consecutive jumps,
first $\tau_1$ and then $\tau_2$. For the deterministic part, the two
jumps add up to the single large jump, so, as shown in
Section~\ref{sec:composition}, the map is approximately consistent with this
composition. For the fluctuations, the total noise of the large jump is
the sum of the noises of the two smaller jumps. If these were completely
independent, their variances would simply add, and the variance would
grow linearly with the separation. The fitted scaling grows slightly more
slowly than linearly, so the two consecutive jumps cannot be exactly
independent: they would need to be very weakly anti-correlated, meaning
that an unusually large jump tends to be followed by a slightly smaller
one. Assuming self-similar fluctuations with a Hurst-like exponent $H$,
the correlation between two consecutive jumps of equal size is
$\rho=2^{2H-1}-1$, which for $H\simeq0.46$ gives $\rho\simeq-0.05$,
i.e., a very weak anti-correlation. Since this value is very small, the
composition of consecutive jumps is compatible with the noise scaling of
the model, at least at the level of the variance. It should be noted that
this correlation follows directly from the power-law scaling and is not
an independent test of the model, which would require measuring the
correlation between consecutive jumps directly from the data.


The iterative map describes the relation between successive primes, but it cannot be used to generate the exact prime sequence, since $s$ is characterized here only by its statistical properties and not by its specific realization at each $n$. While one could in principle sample $s$ from its Gamma distribution to generate a statistically representative surrogate sequence, this would not reproduce the true, deterministic sequence of primes; $s$ remains an unknown stochastic variable, and Eq.~\eqref{eq:full_model} should be understood as a statistical, rather than a deterministic, generative model of the prime sequence.

\section*{Discussion and perspectives}

In this work, prime gaps across multiple separation scales were
analyzed empirically. An iterative map describing the average growth of
successive primes was derived from a modified local approximation of
the Prime Number Theorem. The proposed form of the coefficient $c = \tfrac{1}{2}\ln p_n$ can be interpreted as representing half of the average asymptotic prime gap
over a separation $\tau$, i.e., the distance from $p_n$ to the
approximate midpoint of the interval $[p_n, p_{n+\tau}]$. The
numerical analysis, which minimizes the mean residual fluctuations,
supports the hypothesis that introducing the parameter $c$ into the
equation provides a more accurate description of the leading prime
growth. Notably, the value of $c$ obtained independently from the
theoretical derivation of the iterative map, based on the local Prime
Number Theorem approximation, is consistent with the empirically
optimized value, indicating that this improvement is grounded in the
underlying asymptotic prime distribution rather than being merely a
fitting artifact.

The decomposition of the map into a regular asymptotic transport and a
fine residual component suggests a connection with dynamical systems. In
this interpretation, the dominant transport generates long regular
phases, whereas the small residual terms encode localized irregular
departures from this transport. The residual fluctuations $s=D_{n,\tau}(r-1)$ scale with the deterministic gap $D_{n,\tau}$ and are described by a unit-mean Gamma-distributed factor $r$, so that their distribution is a weighted sum of rescaled Gamma distributions. The relative variance of $r$ follows an approximate power law in the separation distance between primes, and the variance of $s$ therefore does as well.

In this study, the linear differences of non-consecutive primes have been considered. However, one could instead consider differences of specific nonlinear functions of non-consecutive primes, such that part of the residual fluctuations could be absorbed into the nonlinear transformation. Formally, this amounts to asking whether the residual
term $s$ can be expressed as a coboundary of the map, i.e., whether there exists a function $\phi$ such that $\phi(p_{n+\tau}) - \phi(p_n)$ absorbs part of $s$, in analogy with the cohomological equations studied in dynamical systems theory. An interesting question is therefore whether such a representation in nonlinear coordinates could
reveal additional deterministic structure in the residuals $s$, providing an indication that the irregularities in prime gaps could be map-generated. This hypothesis will be the subject of future study.






\section*{Acknowledgements}

The author wish to thank F. Marino for valuable comments on the manuscript.

\end{document}